\documentclass[aip,rsi,amsmath, amsymb,reprint,graphicx]{revtex4-1} 

\usepackage{graphicx}
\usepackage{dcolumn}
\usepackage{bm}
\usepackage{textcomp}

\begin{document}

\title{Dispersion Calibration for the National Ignition Facility Electron Positron Proton Spectrometers for Intense Laser Matter Interactions} 
\author{Jens \surname{von der Linden}}
\email[]{jens.von.der.linden@ipp.mpg.de}
\altaffiliation[Current address: ]{Max Planck Institute for Plasma Physics, 17491 Greifswald and 85748 Garching, Germany}
\affiliation{Lawrence Livermore National Laboratory, Livermore, CA 94550, USA}
\author{Jos\'{e} \surname{Ramos Mendez}}
\author{Bruce Faddegon}
\affiliation{Radiation Oncology, University of California, San Francisco, CA 94143, USA}
\author{Devan Massin}
\affiliation{Applied Physics and Applied Mathematics, Columbia University, New York, NY 10027, USA \looseness=-1}
\author{Gennady Fiksel}
\affiliation{Center for Ultrafast Optical Science, University of Michigan, Ann Arbor, MI 48109, USA \looseness=-1}
\author{Joe P. Holder}
\affiliation{Lawrence Livermore National Laboratory, Livermore, CA 94550, USA}
\author{Louise Willingale}
\affiliation{Center for Ultrafast Optical Science, University of Michigan, Ann Arbor, MI 48109, USA \looseness=-1}
\author{Jonathan Peebles}
\affiliation{Laboratory for Laser Energetics, University of Rochester, Rochester, NY 14623, USA \looseness=-1}
\author{Matthew R. Edwards}
\author{Hui Chen}
\affiliation{Lawrence Livermore National Laboratory, Livermore, CA 94550, USA}

\date{\today}

\begin{abstract}
Electron-positron pairs, produced in intense laser-solid interactions, are diagnosed using magnetic spectrometers with image plates, such as the National Ignition Facility (NIF) Electron Positron Proton Spectrometers (EPPS).
Although modeling can help infer the quantitative value, the accuracy of the models needs to be verified to ensure measurement quality.
The dispersion of low-energy electrons and positrons may be affected by fringe magnetic fields near the entrance of the EPPS.
We have calibrated the EPPS with six electron beams from a Siemens Oncor linear accelerator (linac) ranging in energy from $2.7$--$15.2$ $\mathrm{MeV}$ as they enter the spectrometer.
A Geant4 TOPAS Monte-Carlo simulation was set up to match depth dose curves and lateral profiles measured in water at $100$ $\mathrm{cm}$ source-surface distance.
An accurate relationship was established between the bending magnet current setting and the energy of the electron beam at the exit window.
The simulations and measurements were used to determine the energy distributions of the six electron beams at the EPPS slit.
Analysis of the scanned image plates together with the determined energy distribution arriving in the spectrometer provide improved dispersion curves for the EPPS.
\end{abstract}

\maketitle

\section{Introduction}
Intense laser-matter interactions can produce short bursts of hot electrons, positrons, protons and heavy ions \cite{Gibbon2005} for a variety of potential applications ranging from fast ignition concepts \cite{Tabak2005POP, Roth2001PRL} for inertial fusion energy to diagnostic proton beams \cite{Borghesi2001PPCF} and anti-matter sources \cite{Chen2015PRL}.
Over the last decades progress has been made in understanding the physics generating these particle bursts.
On the front of the target the pre-pulse of the laser produces a preformed plasma \cite{MacPhee2010PRL} and the laser-matter interaction accelerates electrons through  $J \times B$ forces \cite{Wilks1992PRL} and stochastic processes \cite{Kemp2014NF}.
As the relativistic electrons propagate through the target they produce bremsstrahlung and  positrons \cite{Heitler1954}.
On the backside, positrons, protons, and heavy ions experience target-normal sheath acceleration (TNSA) \cite{Wilks2001POP}.
Based on this understanding it is now becoming possible to control these processes in order to optimize the particle bursts for specific applications.
For example, the high numbers ($10^{12}$) of energetic ($>10$ $\mathrm{MeV}$) electrons and positron produced by laser-matter interactions represent an attractive source for pair plasma trapping experiments \cite{Stoneking2020JPP}; however, existing pulsed-power driven magnetic coils\cite{Fiksel2015RSI} at short-pulse laser facilities achieve magnetic fields that can confine pairs of only a few $\mathrm{MeV}$.
To lower the energy of laser-matter interaction generated pairs the sheath on the back side of the target needs to be manipulated to reduce the TNSA.
\par
Progress in understanding and ability to manipulate the processes of intense laser-matter interaction relies on detailed measurements of the particle beams.
Particles can be recorded with image plates \cite{Ohuchi2000NIMPR}, or through imaging of scintillating screens \cite{Glinec2006RSI}.
The spatial distribution of particles and photons can be directly recorded \cite{Rusby2015JPP, Nakamura2011PRSTAB, Buck2010RSI} or particle spectrometers can be employed to separate charged particles by mass and energy with magnetic, or also with electric fields \cite{Alejo2016RSI}.
By using image plates and permanent magnets \cite{Chen2008bRSI, Liang2015SR, Habara2019RSI}, imaging can be done hours later in an external scanner and the magnetic spectrometers can be ruggedized against electromagnetic pulses of laser-matter interactions.
Particle spectrometers can be shielded against, radiation only accepting particles through a narrow slit.
They can provide large energy resolution and, if used in groups, spatial and angular resolution.
\par
Here, we focus on the calibration of a specific family of magnetic spectrometers called electron-positron-proton spectrometers (EPPS) \cite{Chen2008bRSI} developed at Lawrence Livermore National Laboratory (LLNL) specifically for diagnosing intense-laser matter interaction and used at several laser facilities including the National Ignition Facility (NIF) as NIF EPPS (NEPPS).
The EPPS spectrometer uses (two $51$x$102$x$1.3$ $\mathrm{cm}$ Neodymium) permanent magnets to separate charged particles with an uniform magnetic field strength onto BAS-SR image plates on curved holders.
The magnetic field strength and image plate size and shape determine the EPPS energy coverage.
The curved image plates allow the EPPS to resolve electron and positron energies over four orders of magnitude, with high resolving power.
The EPPS image plate scans are interpreted with models of the dispersion of particle energies in the magnetic field to determine the energy of the observed signals, the resolving power, and uncertainties.      
\section{Dispersion models}
The motion of a charged particle is described by the relativistic equation of motion, $\frac{\partial}{\partial t} (\gamma m \vec{u}) = q \vec{u} \times \vec{B}$, where $\gamma$ is the relativistic correction factor, $m$ is the particle mass, $q$ the particle charge, $\vec{u}$ the particle velocity and $\vec{B}$ the magnetic field.
\par
\begin{figure*}
\includegraphics[width=\textwidth,keepaspectratio]{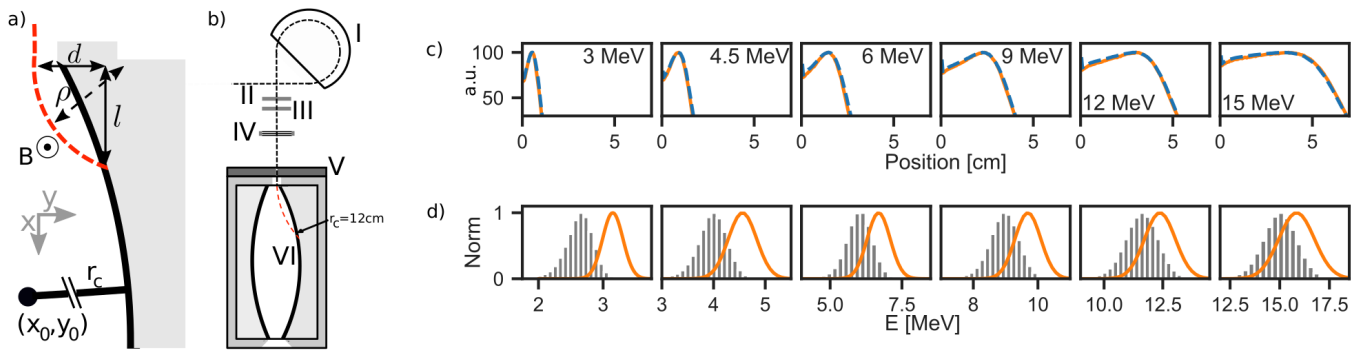}
\caption{\label{fig:setup}
EPPS image plate geometry (a), setup of the calibration measurement (b) (not to scale), standard measurement (c), and source and EPPS entry energy distributions (d).
In (a) an electron (path in red) is deflected by the magnetic field B onto a gyro orbit of radius $\rho$.
The impact position on the image plate is defined by lengths $d$ and $l$.
The image plate is a section of a circle defined by the radius of curvature $r_c$ and the center $(x_0$, $y_0)$.
In (b) the energy of the electron beam is selected by the bending magnet field (I).
The beam passes through an exit window (II), a scattering foil (III), and a monitor chamber (IV).
Upon entering the EPPS through a slit in a tantalum plate (V) the electrons are deflected by the magnetic field B towards the radius of curvature $r_C=12$ $\mathrm{cm}$ image plate (VI).
c) Depth dose curves calculated with Monte-Carlo simulation (orange) of each beam compared to diode measurements (blue dashed) in the water tank.
d) Normalized energy distribution of the source in the simulation (orange) and simulated energy distribution of electrons arriving at EPPS slit (gray bars).}
\end{figure*}
The EPPS is designed to have a uniform magnetic field so the particles will move along the gyroradius of their energy $\rho= \gamma m u_\perp / (q B)$, where $u_\perp$ is the velocity perpendicular to the magnetic field.
Assuming the field is perfectly perpendicular to the plane of the image plate radii, the relationship between gyroradius $\rho$, the displacement towards the image plate $d$ and the displacement along the dispersion direction $l$ (Fig.~\ref{fig:setup} a) can be expressed as,
\begin{equation}
\label{eq:ana_flat_plates}
\rho^2 = \frac{l^2 + d^2}{2 d}.
\end{equation}
In the EPPS the image plates are curved following the circumference of a circle centered at $x_0$ and $y_0$ with radius $r_C$ making l and d interrelated along the image plate,
\begin{equation}
\label{eq:circ_plates}
d =y_0 + \sqrt{r_c^2 - (l - x_0)^2}.
\end{equation}
With these assumptions the dispersion can be expressed as a function of particle energy by combining eqs. (\ref{eq:ana_flat_plates}) and (\ref{eq:circ_plates}), solving for $l$ in terms of arc-length.
Uncertainties occur due to the initial entry position and angle of the particles, both are constrained by the narrow collimating slit.
While the EPPS spectrometers are designed to have less than $10 \%$ variation in magnetic field strength across the dispersion plane \cite{Chen2008bRSI}, at the edges of the plane near the entrance slit there will be 3D fringe fields \cite{Cha2012RSI} and variations in the rise of the magnetic field.
To account for this non-uniformity, the field of the permanent magnets can be calculated \cite{Glenn2019JI} from their magnetization $\vec{M}$ and the divergence-less magnetic field condition $\nabla \cdot (\vec{H} + \vec{M}) = 0$.
In this study we used the COMSOL Multiphysics\textregistered AC/DC solver \cite{COMSOL} to calculate the magnetic field from the EPPS geometry, magnetization of the magnets and permeability of materials.
While the calculated field is mainly unidirectional and uniform throughout most of the EPPS body near the slit there are indeed 3D fringe fields that especially affect the paths of low energy particles.
The uniform field and numerical model can be adjusted to measurements of the peak magnetic field and the magnetic field profile along the slit axis measured by Hall probes, respectively.
However, it is difficult to access the full spectrometer volume through the narrow slit for a complete magnetic field map. 
Verifying how important fringe field effects are and how accurately these are captured by the uniform and numerically calculated field dispersion models requires calibration measurements. 
\section{Calibration}
Spectrometers are calibrated with particle sources such as electron guns \cite{Chen2006RSI}, radio frequency electron accelerators \cite{Zeil2010RSI, Tanaka2005RSI}, or even laser-matter interactions that are characterized with dosimeters \cite{Chen2008aRSI} or filtered with film pack \cite{Mariscal2018RSI}.
Medical electron accelerators are widely used to deposit energy in tumors sufficient to destroy them without damaging surrounding cells.
The detailed modeling and stability of beam characteristics required to achieve successful treatment make electron beams from these accelerators excellent calibration sources \cite{Liang2015SR, Gobet2016RM, Ozaki2014RSI}.
We calibrate the EPPS against a standard measurement for a Oncor medical electron accelerator.
\par
The Oncor medical linac (Siemens Oncology Care Systems, Concord, USA) accelerates electrons from a diode gun in a RF waveguide.
A bending magnet (Fig.~\ref{fig:setup} b) rotates the electron beam direction by $270^\circ$ \cite{Oline1990EPAC:EPAC1990}.
Since the gyroradius depends on the electron energy and magnetic field, setting the field strength or rather the current of the electromagnet selects the beam energy.  
For medical purposes the lowest default beam energy is nominally $6$ $\mathrm{MeV}$; however, for these calibration measurements we have reduced the bending magnet current to achieve energies of $3.15$ $\mathrm{MeV}$ and $4.56$ $\mathrm{MeV}$ at the exit window.
After the bending magnet the electron beam travels through the treatment head \cite{Faddegon2009MP} with components for shaping the beam.
For the calibration measurements we only kept difficult-to-remove components: the water cooled titanium exit window, an aluminum scattering foil, and an electron monitor chamber consisting of several kapton and gold layers (Fig.~\ref{fig:setup} b).
Ref. [\onlinecite{Faddegon2009MP}] characterized the geometry of each of these components.
\par
The standard measurement for this calibration is a depth scan of the beam energy deposited in water.
Two CC13 ionization chambers (IBA Dosimetry) measured the percentage depth profiles in the water tank placed below the treatment head at the accelerator iso-center (100 cm from exit window) \cite{Gobet2016RM}.
One chamber is placed above the water and acts as a reference to account for fluence irregularities, while the other chamber measures the absorbed dose at stepped depths.
To center the chambers initially the measurement chamber scanned two perpendicular lateral profiles. 
For calibration measurements each EPPS is placed below the treatment head ($56.2$ $\mathrm{cm}$ below the exit window) (Fig.~\ref{fig:setup}).
Image plates are placed in the EPPS and exposed to one or several of the beam energies.
After exposure the image plates are stored in light-tight bags and kept in a gel-pack-cooled box to keep the temperature within a consistent range ($13^\circ \mathrm{C}$ -- $18^\circ \mathrm{C}$) until scanning on the next day.
The exposure and scan time are noted to account for image plate fading \cite{Izumi2006RSI}.
\par
To determine the energy of the electron beam arriving at the EPPS slit, the beam source has to be characterized and the scattering in the treatment head components and air taken into account.
This is achieved with Monte-Carlo simulations of the propagation of the electron beams. 
The Monte-Carlo simulations were performed with TOPAS \cite{Perl2012MP}, a wrapper of Geant4 developed in the medical physics community.
TOPAS includes Geant4 physics lists describing multiple electron scattering.
The geometry and materials of treatment head components are taken from the characterizations of a previous study \cite{Faddegon2009MP}.
The electron beam is assumed to have a Gaussian energy profile and simulated with $5\cdot 10^7$ particle histories to provide sufficient statistical precision in the calculated quantities.
The mean and width of the Gaussian-shaped source distribution are adjusted until simulated depth dose in water agrees with those of the standard measurements (Fig.~\ref{fig:setup}c).
Then the beam is scored at the height of the EPPS slit.
The source energy distribution and energy distribution arriving at the EPPS slit height are plotted for each nominal beam energy in fig.~\ref{fig:setup}d.
The bending magnet voltage and beam parameters are summarized in table \ref{tab:table_beams}.
The bending magnet voltage is listed and the impedance relating the bending magnetic voltage and current is $6.19 \pm 0.05$ $\mathrm{\Omega}$.
The bending magnet voltage $V_B$ and the mean energy of the electron beam at the exit window $\mu_E$ fit a line given by $\mu_E = 3.33$ $[\mathrm{MeV/V}]$ $V_B - 0.353$ $[\mathrm{MeV}]$ within $0.3 \%$.
This linear relationship also holds for the two beam energies below the standard operating regime.
The simulated beam energy distribution parameters and angles at the EPPS slit height are used for the following comparison to the measured dispersion positions of the beams in the EPPS. 
Further scattering will occur along the electron path of the air filled EPPS but this effect is small since a $1$ $\mathrm{MeV}$ electron loses about $2$ $\mathrm{keV}$ per cm in air at room temperature and pressure and the combined length of EPPS slit material and dispersion volume is $\sim 11$ $\mathrm{cm}$.
\begin{table}
\caption{\label{tab:table_beams} Beam parameters for each nominal beam energy $E_N$: Bending magnet voltage $V_B$, mean energy at exit window $\mu_E$, mean energy at EPPS slit $\mu_S$, standard deviation in energy at slit $\sigma_S$, and standard deviation in angle $\sigma_\circ$}
\begin{ruledtabular}
\begin{tabular}{cccccc}
  $E_N$ [$\mathrm{MeV}$] &
  $V_{B}$ [$\mathrm{V}$] &
  $\mu_{E}$ [$\mathrm{MeV}$] &
  $\mu_S$ [$\mathrm{MeV}$] &
  $\sigma_S$ [$\mathrm{MeV}$] &
  $\sigma_\circ$ [$\mathrm{deg}$] \\
\hline
3 & 1.050 & 3.15 & 2.69 & 0.19 & 0.1 \\

4.5 & 1.475 & 4.56 & 4.04 & 0.26 & 0.08 \\

6 & 2.110 & 6.69 & 6.13 & 0.42 & 0.07 \\

9 & 3.018 & 9.69 & 9.06 & 0.43 & 0.06 \\

12 & 3.819 & 12.38 & 11.73 & 0.67 & 0.05 \\

15 & 4.860 & 15.85 & 15.17 & 0.82 & 0.04 \\
\end{tabular}
\end{ruledtabular}
\end{table}
\begin{figure}
\includegraphics[width=\columnwidth,keepaspectratio]{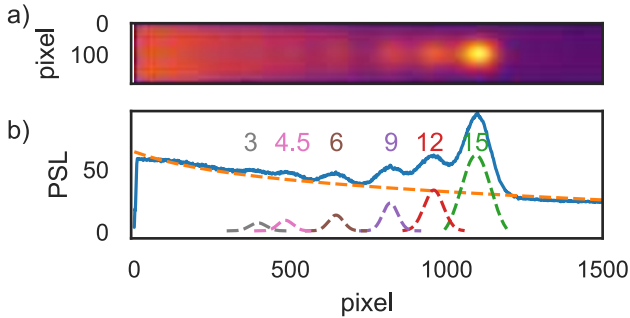}
\caption{\label{fig:data_analysis} Scan of image plate exposed to all six beams a) and fits of the spots produced by the beams b). In b) a horizontal line-out of the scan (solid blue), a double exponential fit to the background (dashed orange) and six Gaussian fits to the beam spots (dashed) with the nominal beam energy labels are plotted.}
\end{figure}
57 calibration image plates were exposed to electron beams in six EPPS models. 
Each shot exposed an image plate to between one and all six of the available electron beam energies.
Fig.~\ref{fig:data_analysis}a) shows an image plate exposed to all six energies. 
The image plate scans show spots that correspond to the electron beams on top of a background.
The background signal has a super-Gaussian shape in the vertical direction \cite{Park2015} and a double-exponential shape in the horizontal direction, increasing in magnitude towards the low energy side \cite{Williams2016}.   
The beam spots are super-Gaussian in the vertical direction and Gaussian in the horizontal direction.
The shape in the horizontal direction results from the energy distribution of the electron beam and the dispersion of equal energy electrons entering across the finite slit width ($\sim 1$ $\mathrm{mm}$).
To determine the position of the beam spots a double exponential is fit to the background and subtracted from the signal, then Gaussians are fit to an averaged line-out of the signal (Fig.~\ref{fig:data_analysis} b).
The $3$ $\mathrm{MeV}$ and $4.5$ $\mathrm{MeV}$ beam spots have low signal-to-noise ratios on image plates exposed to all beams.
On image plates solely exposed to the $3$ or $4.5$ $\mathrm{MeV}$ beams the signal-to-noise ratio is higher, $\sim2$ ($\sim10$ with background fit subtraction), because the background is related to the total dosage.  
\begin{figure}
\includegraphics[width=\columnwidth,keepaspectratio]{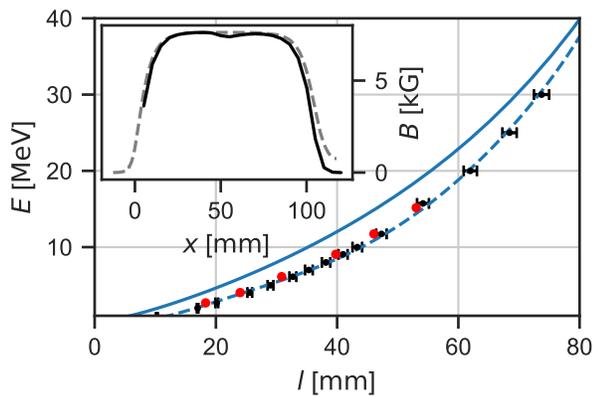}
\caption{\label{fig:dispersion} Average peak position of the beam spots for each energy plotted (red) together with the dispersion models using a $0.75$ $\mathrm{T}$ uniform magnetic field (solid blue), and a numerically calculated magnetic field (black markers with dashed interpolating polynomial), which includes fringe effects.
The error bars correspond to the dispersion due to simulated electrons entering across the finite slit width.
$d=0$ is the start of the image plate.
Inset shows the magnetic measured (solid black) with a Hall probe along the central axis compared to the calculated magnetic field (dashed gray).
$x=0$ is the slit position.}
\end{figure}
Here, we present the calibration of the EPPS4-H, a specific EPPS, in which 15 image plates were exposed.
In the EPPS-4H the magnetic field rises over the first $20$ $\mathrm{cm}$ of the main axis, reaches a $0.75$ $\mathrm{T}$ magnitude with variations of only $6\%$ along the central $60$ $\mathrm{cm}$ (fig.~\ref{fig:dispersion} inset).
For each beam energy there are 3--4 exposed image plates.
Based on previous sensitivity analysis of the treatment head model the simulated beam energies are accurate to within $1\%$ \cite{Faddegon2009MP}. 
The standard deviation of the peaks of the beam spots ranged from $800$ $\mathrm{\mu m}$ (16x$50$ $\mathrm{\mu m}$ pixels) for the $3$ $\mathrm{MeV}$ beam to $200$ $\mathrm{\mu m}$ (4 pixels) for the $9$ and $12$ $\mathrm{MeV}$ beams.
Fig.~\ref{fig:dispersion} shows the average peak position $l$ of the beam spots for each energy $E$ plotted (red) together with the dispersion models of uniform magnetic field (solid blue), and numerically calculated magnetic field (black markers).
The $l$ of the uniform field dispersion model differs by $12 \%$--$29 \%$ from the calibration beam spots.
As the energy increases the uniform field model approaches the numerical dispersion.
Higher energy electrons are less affected by the non-uniformity in the magnetic field near the slit.
The $l$ of the numerically calculated (uniform field) dispersion model differs by $10 \%$--$2 \%$ ($12 \%$--$29 \%$) from the calibration $l$.
The resolving power $dE/dl$ can be determined by differentiating a interpolating 6th polynomial given by $l = \sum_{n=0}^{n=6} c_n \cdot E^n$, where $c_0=1.29$, $c_1=-0.419$, $c_2=5.11\cdot 10^{-2}$, $c_3=-1.95\cdot 10^{-3}$, $c_4=2.83\cdot 10^{-5}$, $c_5=-3.56\cdot 10^{-7}$, $c_6=1.3\cdot 10^{-9}$, $l$ [$\mathrm{mm}$], and $E$ [$\mathrm{MeV}$].
Multiplying the resolving power by the difference in $l$ of the calibration and the numerical dispersion model yields a $<0.7$ $MeV$ error for all calibration beams.
\section{Conclusions}
The calibration measurements with six well-characterized electron beams confirm that a dispersion model based on numerically calculated magnetic fields based on spectrometer geometry and magnetization values provides an accurate description of the dispersion.
The resulting dispersion curve agrees with the measurements within $<0.7$ $\mathrm{MeV}$ over the energy range of the calibration measurements, $2.7$--$15.2$ $\mathrm{MeV}$.
The error of the dispersion calculated from a uniform field is $\sim 3$--$6$ times the error dispersion from a numerically calculated field.
At higher energies the effect of fringe fields is diminished and the numerical dispersion approaches the dispersion of electrons in a uniform magnetic field.
The image plate scans from these calibration measurements could be used to calibrate the dosage response of image plates in the EPPS to the fluence of an electron beam as determined by ion chamber dosimetry\cite{TG-51}.
A future study could test whether the background signal observed in these calibration measurements and laser-matter interaction experiments originates from bremstrahlung by further propagating electrons in a Monte-Carlo simulation through the full spectrometer geometry \cite{Gobet2016RM}, tracking their scattering off the EPPS material, and scoring the resulting photons at the image plates.  

\section*{Acknowledgments}
This work was performed under the auspices of the U.S. Department of Energy by Lawrence Livermore National Laboratory under Contract DE-AC52-07NA27344 and was supported by the LLNL-LDRD Program under Project No. 20-LW-021.
We acknowledge E. Grade for assistance operating the linac, N. Izumi for discussing handling of image plates, and M. S. Beach for providing a temperature-controlled box. LLNL-JRNL-817275.

\section*{Data Availability}
The data that support the findings of this study are available from the corresponding author upon reasonable request.

\end{document}